\documentclass[a4paper, amsfonts, amssymb, amsmath, reprint, showkeys, nofootinbib, twoside, superscriptaddress, floatfix]{revtex4-1}
\usepackage{natbib}
\usepackage[pdftex, pdftitle={Article}, pdfauthor={Author}]{hyperref} 
\usepackage{amsmath,amssymb,amsfonts}
\usepackage{algorithmic}
\usepackage{graphicx}
\usepackage{bm}
\usepackage[normalem]{ulem} 
\usepackage{multirow} 
\usepackage{booktabs} 
\usepackage{ulem} 
\usepackage{gensymb}
\usepackage[utf8]{inputenc}
\usepackage{textcomp}
\usepackage{siunitx}
\usepackage{soul} 
\usepackage[normalem]{ulem} 
\usepackage{multirow} 
\usepackage{booktabs} 
\usepackage{ulem} 
\usepackage[switch]{lineno}
\soulregister\cite7
\soulregister\ref7
\soulregister\eqref7

\newcommand{\mb}[1]{\mathbf{#1}}
\newcommand{\mbx}{\mb{x}}

\newcommand{\dd}{\mathrm{d}}

\begin{document}
\title{\texorpdfstring{Ultrasound Autofocusing:\\ Common Midpoint Phase Error Optimization via Differentiable Beamforming}{Ultrasound Autofocusing: Common Midpoint Phase Error Optimization via Differentiable Beamforming}}

\author{Walter Simson}
\thanks{Equal contribution.}
\affiliation{Department of Radiology, School of Medicine, Stanford University, Stanford, CA 94305 USA}

\author{Louise Zhuang}
\thanks{Equal contribution.}
\affiliation{Department of Electrical Engineering, Stanford University, Stanford, CA 94305 USA}

\author{Benjamin N. Frey}
\affiliation{Department of Applied Physics, Stanford University, Stanford, CA 94305 USA}

\author{Sergio J. Sanabria}
\affiliation{Department of Radiology, School of Medicine, Stanford University, Stanford, CA 94305 USA}
\affiliation{Ikerbasque, Basque Foundation for Science, Bilbao, Spain}

\author{Jeremy J. Dahl}
    \email[Correspondence email address: ]{jjdahl [at] stanford [dot] edu}
\affiliation{Department of Radiology, School of Medicine, Stanford University, Stanford, CA 94305 USA}

\author{Dongwoon Hyun}
\affiliation{Department of Radiology, School of Medicine, Stanford University, Stanford, CA 94305 USA}
\affiliation{Siemens Healthineers, Palo Alto, CA 94304 USA}

\date{August 15, 2025}


\setstcolor{red}


\begin{abstract}
In ultrasound imaging, propagation of an acoustic wavefront through heterogeneous media causes phase aberrations that degrade the coherence of the reflected wavefront, leading to reduced image resolution and contrast. Adaptive imaging techniques attempt to correct this phase aberration and restore coherence, leading to improved focusing of the image. We propose an autofocusing paradigm for aberration correction in ultrasound imaging by fitting an acoustic velocity field to pressure measurements, via optimization of the common midpoint phase error (CMPE), using a straight-ray wave propagation model for beamforming in diffusely scattering media. We show that CMPE induced by heterogeneous acoustic velocity is a robust measure of phase aberration that can be used for acoustic autofocusing.  CMPE is optimized iteratively using a differentiable beamforming approach to simultaneously improve the image focus while estimating the acoustic velocity field of the interrogated medium. The approach relies solely on wavefield measurements using a straight-ray integral solution of the two-way time-of-flight without explicit numerical time-stepping models of wave propagation. We demonstrate method performance through \emph{in silico} simulations, \emph{in vitro} phantom measurements, and \emph{in vivo} mammalian models, showing practical applications in distributed aberration quantification, correction, and velocity estimation for medical ultrasound autofocusing.
\end{abstract}


\maketitle

\section{Introduction}
\label{sec:introduction}\label{sec1}
Coherent imaging systems reconstruct the physical properties of distant sources using phase-sensitive wavefield measurements, such as radio waves in radar, visible light in optics, or acoustic waves in sonar, seismic, and ultrasound imaging. The search for signal coherence given an underlying propagation model is the unifying principle that underpins these modalities. Discrepancies between an assumed phase propagation model and the true propagation result in phase aberration, a defocusing effect that degrades the resolution and contrast of a coherent imaging system. For ultrasound imaging, phase aberration results from unmodeled tissue velocity in the propagation medium and produces degraded images~\cite{ali2023aberration}.

Phase screen models have been adopted in ultrasound imaging, where phase (or time) shifts between neighboring aperture elements are measured and iteratively corrected to compensate for phase aberration \cite{flax1988phase, Odonnell1988PhaseAb}. Although phase screens are reasonable models of thin aberrating media near the aperture, such as a thin layer of subcutaneous fat, they do not accurately model distributed aberration encountered in heterogeneous media, such as the breast or thick layers of subcutaneous tissue \cite{wu1992timereversal,liu1994correction,hinkelman1998abdominal}. Wave propagation through heterogeneous biological tissue with varying acoustic velocities results in spatially dependent phase aberration that accumulates over the wave's travel path, which is also known as distributed phase aberration. Distributed phase aberration correction can be achieved by modifying the wave propagation model to better match the true propagation.

Acoustic full waveform inversion (FWI) fits the wave equation operator and initial conditions to a template of wavefield measurements to recover physical domain properties such as absorption, acoustic velocity, or density. However, FWI is ill-posed for non-smooth property distributions, non-convex, computationally intensive \cite{virieux2009overview}, and does not take advantage of far-field coherence patterns to regularize the inverse-problem. Alternatively, seismic focusing analysis imaging techniques such as migration velocity analysis (MVA) and wave equation migration velocity analysis (WEMVA) update only the wavefield velocity used in a propagation model to optimize the flatness of specular reflectors in seismic images \cite{biondi20063d,sava2004wave}. While MVA uses a straight-ray operator and WEMVA uses a linearized wave equation operator, both methods allow for reflection tomographic reconstruction of wave velocity given well-defined specular reflections. In medical ultrasound where diffuse scattering is prevalent, reflection tomographic methods \cite{jaeger2015computed,sanabria2018spatial,stahli2020improved,bezek2023analytical,ali2022distributed,ali2023sound} have been proposed to estimate spatially varying velocity distributions based on matrix inversion of a forward phase-shift distribution model. Using recovered velocity distributions, phase aberration correction in ultrasound imaging has been achieved using straight ray, Eikonal, and linear wave propagation models \cite{jaeger2015correction,ali2023aberration,rau2019ultrasound,ali2022distributed}.

In autofocusing, phase aberration correction is applied iteratively to improve focus quality. A reliable focusing measure with the following characteristics is required to perform ultrasound autofocusing: it is suitable for diffuse targets, has a well-defined optimal value, and has minimal phase randomness, i.e., jitter. Many existing focusing criteria for pulse-echo ultrasound do not display these properties. The classic main lobe width of the point spread function (PSF) describes the imaging system resolution but traditionally requires a point source in the field of view, which requires manual target selection and is not possible in diffuse scattering. Speckle brightness \cite{nock1989phase}, defined as the coherent sum of a diffuse scatterer source, is maximized when ideally focused, but the optimal value is target-dependent and unbounded. The coherence factor (CF)~\cite{mallart1994adaptive,hollman1999coherence}, defined as the magnitude of the speckle brightness normalized by the incoherent signal sum, is bounded between 0 and 1, but the optimum value is also target-dependent. The optimal CF value under ideal focusing for incoherent sources (diffuse scattering) is 2/3, while the optimal CF value for coherent sources is 1. The spatial focusing factor ($F$-factor) is the quotient of a measured common midpoint PSF over the ideal common midpoint PSF at a given location in space \cite{lambert2022umi1}. The $F$-factor relies on a model fit of an idealized PSF and Gaussian noise and requires a sufficiently sized isoplanatic patch.
Lastly, phase (or time) shifts between neighboring elements \cite{flax1988phase} can be used on diffuse targets, but contain jitter due to the stochasticity of incoherent sources.

In this work, we introduce common midpoint phase error (CMPE), a focusing measure derived from first principles enabling deterministic focusing in diffusely scattering media. We demonstrate that CMPE is a robust quantitative measure of image focus which can be used to evaluate the spatial distribution of phase aberration. By modeling diffuse scattering as an incoherent source \cite{mallart1991van}, the van Cittert-Zernike (VCZ) theorem \cite{goodman2015statistical,mallart1991van} describes the mutual coherence of the wave front across the aperture. VCZ defines the mutual coherence of an incoherent source across a far-field aperture as the Fourier transform of the spatial distribution of the source intensity. Symmetries in the VCZ theorem allow the modulation term to fall from the coherence equation when measurement points are equidistant from the center axis, thereby maximizing correlation~\cite{goodman2015statistical}. For a pulse-echo ultrasound imaging system, this symmetry requires that transmit and receive aperture pairs share a common midpoint
~\cite{rachlin1990direct,larose2006correlation,ng1997speckle,Li1997parti}.

We build on this symmetry to derive the CMPE, which allows for the direct spatial quantification of phase aberration in diffusely scattering media. We apply CMPE to ultrasound autofocusing and velocity estimation in diffuse media. To this end we optimize velocity distributions using differentiable beamforming, as introduced in~\cite{simson2023dbua}. We further show that autofocusing via CMPE minimization concurrently produces spatially resolved velocity estimates. We base our evaluation on simulations, phantom measurements, and \textit{in vivo} mammalian models. We display the practicality of the proposed CMPE autofocusing and its application in localization and quantification of phase aberration and velocity estimation.

\section{Theory}
\subsection{Pulse-Echo Coherence}
Assuming Fresnel (small angle) and Fraunhofer (far-field) approximations \cite{goodman2005introduction}, the pulse-echo signal due to transmit aperture function $T(\mbx_t)$ in plane $\mathcal{T}$, scattering function $S(\mbx)$ in plane $\Omega$, and receive aperture function $R(\mbx_r)$ in plane $\mathcal{R}$ is: 
\begin{equation}
  P(T,S,R) \approx \frac{e^{j2kz}}{(4\pi k)^2} \int_\Omega \tilde{T}(\mathbf{x}) S(\mathbf{x}) \tilde{R}(\mathbf{x}) e^{j\frac{k}{z}\mbx^\top\mbx} \, d\mathbf{x},
\end{equation}
where $k=2\pi/\lambda$ is the wavenumber, $\lambda$ is the wavelength, $^\top$ is the transpose operator, $\mathbf{x}$ is a 2D spatial position in the scattering plane at axial distance $z$ from the aperture planes, and $\tilde{T}(\mbx)$ and $\tilde{R}(\mbx)$ are spatial 2D Fourier transforms of the transmit and receive apertures (e.g. $\tilde{R}(\mbx) = \int_\mathcal{R} R(\mbx_r) \exp \left[-j\frac{k}{z}\mbx^\top\mbx_r\right] \dd{\mbx_r}$, where $\mbx_r$ is the 2D spatial position in the receive aperture plane). In this Fourier transform, the aperture planes are assumed to be coincident and the spatial frequencies are transformed to spatial coordinates by $k=2\pi\mbx/\lambda z$. Note that this formulation assumes that the transmit and receive functions are in the far-field of the apertures, or equivalently, that the apertures are correctly focused without aberration. 

The mutual coherence \cite{goodman2015statistical} of two pulse-echo signals (i.e. their complex correlation) can be written as
\begin{align}
    \Gamma_{12} &= \mathbb{E}\left[P_1(T_1, S_1, R_1) P_2^*(T_2, S_2, R_2)\right], 
    \label{eq:mutual_coherence}
\end{align}
where $P^*$ denotes the complex conjugate of $P$. In biological tissue, the scattering is commonly modeled as diffuse and spatially incoherent, such that the expected value of the correlation of two signals equals the magnitude of $S_0$ when $\mathbf{x}_1$ equals $\mathbf{x}_2$ and is zero otherwise:
\begin{align}
    \mathbb{E}\left[ S_1(\mbx_1) S_2^*(\mbx_2) \right] = |S_0|^2\delta(\mbx_1-\mbx_2).
    \label{eq:spatial_incoherence}
\end{align}
Under this model, the mutual coherence simplifies to
\begin{align}
    \Gamma_{12}
    &\approx \frac{|S_0|^2}{(4\pi k)^4} \int_\Omega 
    \tilde{T}_1(\mb{x})\tilde{T}_2^*(\mb{x})
    \tilde{R}_1(\mb{x})\tilde{R}_2^*(\mb{x})
    \dd{\mb{x}}.
    \label{eq:vcz}
\end{align}
This result generalizes the van Cittert-Zernike theorem \cite{goodman2015statistical,mallart1991van} to incorporate large receive apertures \cite{walker1997speckle}, and predicts that the coherence of arbitrary transmit-receive pairs $(T_1, R_1)$ and $(T_2, R_2)$ depends only on the aperture functions and not the spatially incoherent scattering function $S$ given a constant sound speed. In the presence of aberration, $\Gamma_{12}$ will contain a complex exponential term due to incorrect focusing.

\subsection{The Common Midpoint Phase Error Measure}
\label{sec:cmpe}
\begin{figure}
    \centering
    \includegraphics[width=\linewidth]{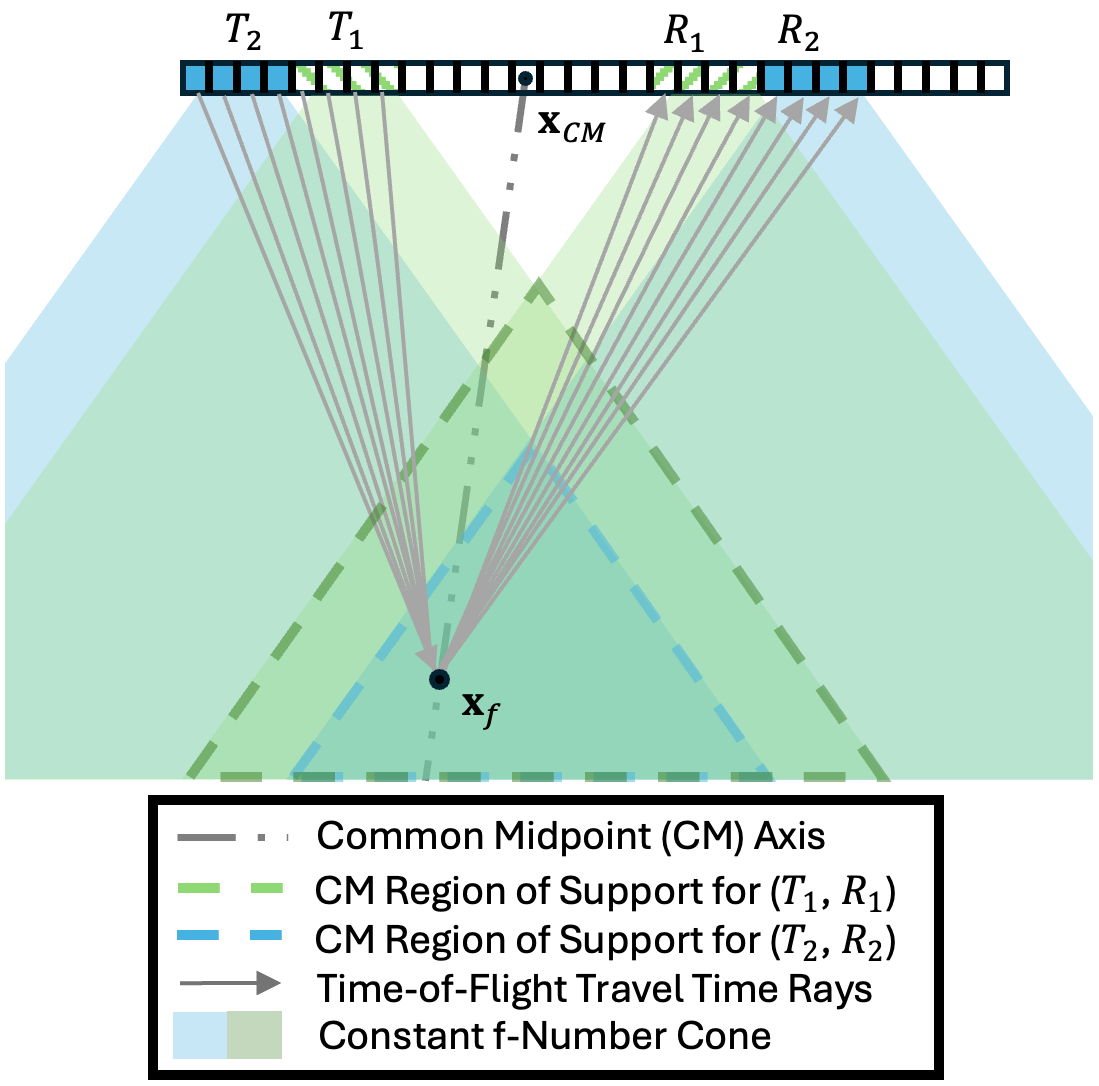}
    \caption{Straight-ray model for two common midpoint subapertures pairs, ($T_1$,$R_1$) and ($T_2$,$R_2$), that are offset by lag-4. The green and blue triangular regions outlined with dashed lines define the region of support, based on a specific $F$/\#, where a common midpoint signal can be observed for the subaperture pair.  The common midpoint $\mbx_{CM}$ defines the possible subaperture pairs, and all common midpoint signals at the reconstruction point $\mbx_f$ in the imaging domain are perfectly correlated.}
    \label{fig:modellayout}
\end{figure}

Consider the case of translating apertures \cite{ng1997speckle}, where $ T_1(\mbx_t+\Delta\mbx_t) = T_2(\mbx_t) = T(\mbx_t)$ and $ R_1(\mbx_r+\Delta\mbx_r)=R_2(\mbx_r) =R(\mbx_r)$. Then, \eqref{eq:vcz} further simplifies as
\begin{align}
    \Gamma_{12}^\textrm{TA} \approx
    \frac{|S_0|^2}{(4\pi k)^4} \int_\Omega 
    |\tilde{T}(\mbx)|^2
    |\tilde{R}(\mbx)|^2 e^{jkz^{-1}\mbx^\top(\Delta\mbx_t+\Delta\mbx_r)}
    \dd{\mbx}.
    \label{eq:translatingapertures}
\end{align}
When $\Delta\mbx_t+\Delta\mbx_r=0$, i.e., when $(T_1,R_1)$ and $(T_2, R_2)$ share a \emph{common midpoint} (CM), the exponential term vanishes:
\begin{align}
    \Gamma_{12}^\textrm{CM} \approx
    \frac{|S_0|^2}{(4\pi k)^4} \int_\Omega 
    |\tilde{T}(\mbx)|^2
    |\tilde{R}(\mbx)|^2 
    \dd{\mbx},
    \label{eq:cmp}
\end{align}
resulting in a real-valued mutual coherence with zero complex angle. Furthermore, the mutual coherence function (essentially a correlation), defined as $\gamma_{12}=\Gamma_{12}/\sqrt{\Gamma_{11}\Gamma_{22}}$, becomes
\begin{align}
    \gamma_{12}^\textrm{CM} = \frac{\frac{|S_0|^2}{(4\pi k)^4} \int_\Omega|\tilde{T}(\mbx)|^2|\tilde{R}(\mbx)|^2\dd{\mbx}}
    {\sqrt{\left[\frac{|S_0|^2}{(4\pi k)^4} \int_\Omega|\tilde{T}(\mbx)|^2|\tilde{R}(\mbx)|^2\dd{\mbx}\right]^2}} = 1.
    \label{eq:gamma12_cmp}
\end{align}

Two focused common midpoint signals are expected to result in zero phase shift ($\angle\Gamma^\textrm{CM}_{12}=0$) and ideal correlation coefficient ($\gamma^\textrm{CM}_{12}=1$), even when the medium is composed of random, diffuse, spatially incoherent scatterers. Conversely, any aberration or model errors will introduce a complex phase term into the mutual coherence function (i.e. $\angle\Gamma^\textrm{CM}_{12}\ne 0$).

Therefore, the CMPE loss is computed as the mean absolute phase shift deviation from zero over all pixels $\mbx_f$ in the field of view (FOV):
\begin{align}
    \mathcal{L}_\textrm{CMPE} = \frac{1}{N_\mbx N_\textrm{CM}}\sum_{\mbx_f\in\textrm{FOV}}\sum_{(p,q)\in\textrm{CM}} |\angle\hat{\Gamma}^\textrm{CM}_{pq}(\mbx_f)|,
    \label{eq:cmpe}
\end{align}
where $\angle\hat{\Gamma}^\textrm{CM}_{pq}(\mbx_f)$ denotes the  estimated phase error between common midpoint signal pairs $(p,q)$ at position $\mbx_f$, $N_\mbx$ is the number of pixels evaluated in the field-of-view, and $N_\textrm{CM}$ is the number of common midpoint pairs evaluated. The CMPE quantifies the aberration and model error present in a given image. The estimation jitter of the phase-shift between any two speckle signals is given by \cite{goodman2020speckle}
\begin{align}
    \sigma^2_{\angle\Gamma}(\rho) &=
     \frac{\pi^2}{3}-\pi\arcsin(\rho) + (\arcsin(\rho))^2 - \frac{1}{2}\textrm{Li}_2(\rho^2),
     \label{eq:jitter}
\end{align}
where $\rho=|\gamma_{12}|$ is the magnitude of the correlation coefficient and $\textrm{Li}_2(\cdot)$ the dilogarithm function. A remarkable consequence of common midpoint signals having ideal correlation ($\rho\approx 1$) is that the CMPE has zero expected jitter, i.e. $ \sigma^2_{\angle\Gamma}(\rho) \approx 0$.

Fig.~\ref{fig:modellayout} illustrates a lag-4 common-midpoint for subaperture pairs $(T_1,R_1)$ and $(T_2, R_2)$, where each subaperture is composed of 4 elements and the two subaperture pairs share a common midpoint at $\mbx_{CM}$. Lag-$n$ refers to the subaperture pairs having a $+n$-element translation between the transmitters of each pair and a $-n$-element translation between the receivers of each pair; in this case, an $n=4$ element translation is described. The point $\mbx_f$ depicts a position where the CMPE can be computed. The two triangular regions outlined by the dashed lines indicate the region of support, subjected to a specific F/\#, where a common midpoint signal can be obtained. Of note is that $\mbx_f$ does not need to be positioned normal to the common-midpoint $\mbx_{CM}$ on the aperture because a ``dynamic near-field correction'' \cite{Li1997parti} is applied via time-delays, allowing $\mbx_f$ to closely approximate a common midpoint signal.

\subsection{Autofocusing with Optimized CMPE via Differentiable Beamforming}

\begin{figure*}
    \centering
    \includegraphics[width=\linewidth]{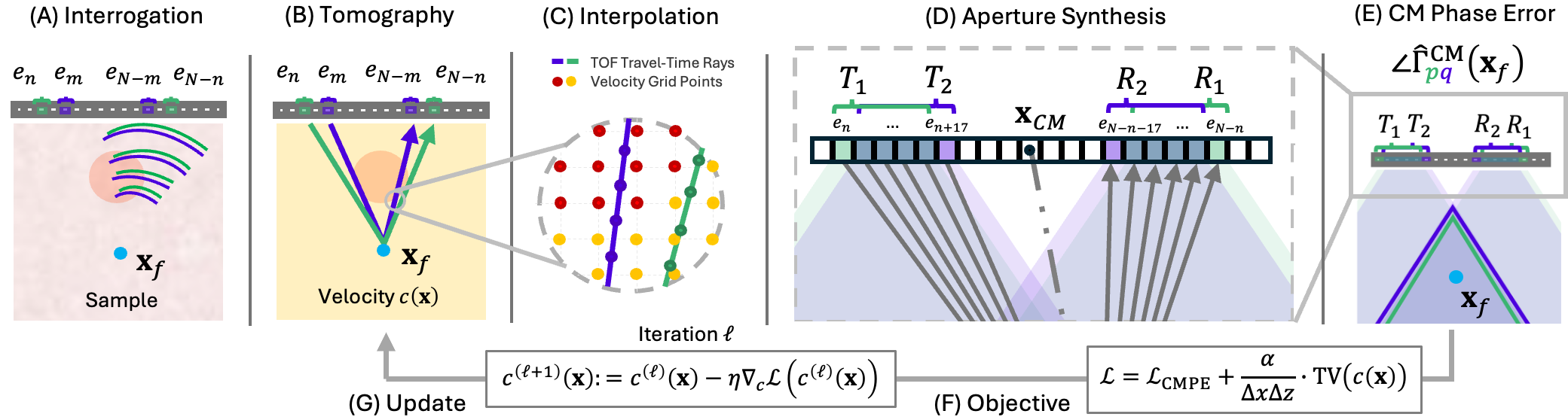}
    \caption{The forward model used for ultrasound autofocusing. (A) Signals from individual elements (e.g. $e_n$, $e_{N-n}$) are captured with a full synthetic aperture sequence. (B) Given a point of interest, $\mbx_f$, a time-of-flight model is derived from the straight-ray reflection tomographic interrogation of an acoustic velocity field. (C) Velocities along the ray are determined via bilinear interpolation of the velocity grid. (D) Larger, possibly overlapping, transmit and receive subapertures, e.g. ($T_1$,$R_1$), are synthesized from the element signals. (E) The CMPE is computed at all points $\mbx$ between common midpoint subaperture signals (green and purple outlines). (F) The objective function, $\mathcal{L}$, is computed and (G) its gradients with respect to the velocity are backpropagated through the forward model via the AMSGrad optimizer to update the velocity at all locations $\mbx$ in (B). The model is iterated ($\ell$) from (B) to (G). Note that, for illustrative purposes only, a conventional gradient descent update with step size $\eta$ is shown in (G).}
    \label{fig:dbuaoverview}
\end{figure*}

An adaptive image reconstruction method for focusing and image correction in diffuse scattering media is formulated using a differentiable beamforming optimization pipeline \cite{simson2023dbua}. A differentiable imaging operator is constructed to focus wavefields from diffuse sources using the computational concept of auto-differentiation \cite{griewank2008evaluating}, which enables an efficient calculation of the adjoint \cite{giles2000introduction,zhu2021general}. The multistatic or full synthetic aperture (FSA) pulse-echo signals are collected as step one of the proposed method (Fig.~\ref{fig:dbuaoverview}A). Focused subaperture signals are then beamformed from transmit and receive elements that share a common midpoint in the outlined regions in Fig.~\ref{fig:modellayout}. 

In beamforming these subaperture signals at $\mbx_f$, a simplified solution to the wave equation (i.e. constant velocity, $c$,  assumption leading to a straight-ray model of wave propagation) is used as an inverse model to assign received signals to their assumed spatial position. Time-of-flight values can be thus calculated as the spatial integral $t_{e\rightarrow f} = \frac{1}{c}\int^{\mbx_{f}}_{\mbx_{e}}\,d\mbx$. In the differentiable beamformer, a reflection tomographic model (Fig.~\ref{fig:dbuaoverview}B) is constructed to model straight-ray propagation in heterogeneous velocity, $c(\mbx)$, between the subapertures, where the time-of-flight values are computed as $ t_{e\rightarrow f}  = \int^{\mbx_{f}}_{\mbx_{e}}\frac{1}{c(\mbx)}\,d\mbx$. Here, the round-trip time-of-flight is computed as the straight-path integral of the slowness, $1/c(\mbx)$, between the transmitter to the focal point and back to the receiver. The slowness function is discretized into a grid and sampled along the rays via bilinear interpolation of the nearest four slowness grid points to the ray's sample location (Fig.~\ref{fig:dbuaoverview}C). 

After calculating a time-of-flight, the element signals are interpolated from the recorded FSA data and assigned to the focal point. 
A set of adjacent transmit elements and adjacent receive elements are then coherently compounding to synthetically form transmit and receive subapertures having common midpoints (Fig.~\ref{fig:dbuaoverview}D) \cite{walker1997speckle}. The focused signal at $\mbx_f$ from the subaperture pairs (e.g. $T_1$ \& $R_1$) is correlated with focused signals from other subaperture pairs sharing a common midpoint (e.g. $T_2$ \& $R_2$) at $\mbx_f$ to estimate the CMPE (Fig.~\ref{fig:dbuaoverview}E). 
Any measured phase error in the correlation of these signals is attributed to beamforming errors along the straight-ray propagation path, and can be integrated into an objective function, $\mathcal{L}$ (Fig.~\ref{fig:dbuaoverview}F), to assess the model. 

Importantly, the forward model and CMPE are formulated as compositions of differentiable operations in JAX \cite{jax2018github}. This allows for calculation of the derivative of the objective function with respect to the velocity via automatic differentiation \cite{zhu2021general}, permitting backpropagation of the total CMPE through the forward model to update the velocity field at each location $\mbx_f$ via the AMSGrad optimizer (a variant of stochastic gradient descent, Fig.~\ref{fig:dbuaoverview}G) 
\cite{rumelhart1986learning,reddi2018amsgrad}. This method seeks the optimal velocity field that minimizes CMPE given the straight-ray forward model, resulting in an autofocused image.

\section{Experimental Methods}
\subsection{Implementation of Differentiable Beamforming}
IQ FSA signals were retrospectively beamformed using $N$ uniformly spaced and overlapping subapertures $\{T_1, T_2, \ldots, T_N\}$ and $\{R_1, R_2, \ldots, R_N\}$. The resulting common midpoint signals formed with the $i$-th transmit and $j$-th receive subaperture is denoted as $P_{i,j}(\mbx)$, and all possible lag-1 transmit-receive subaperture pairs sharing a common midpoint were used to form these common midpoint signals. Subapertures were synthesized with 17 elements each, 
based on empirical observations, to balance the number of available common-midpoint subaperture pairs, accuracy of phase error estimation, and signal-to-noise ratio (SNR). For the implementation of CMPE in \eqref{eq:cmpe}, only lag-1 common-midpoint signals having a correlation exceeding a threshold of 0.7 are used because they minimize phase-wrapping errors and spurious CMPE from noise sources other than aberration. 

To ensure stable and physically plausible velocity distributions, total variation (TV) regularization was applied to the CMPE optimization objective:
\begin{equation}
\mathcal{L} = \mathcal{L}_{\textrm{CMPE}} + \frac{\alpha}{\Delta x\Delta z} \cdot \textrm{TV}\left(c\left(\mbx\right)\right)
\end{equation}
where the total variation term is defined as:
\begin{equation}
\textrm{TV}\left(c\left(\mbx\right)\right) = \sum_{i,j} w_x \left| \frac{\partial c\left(\mbx\right)}{\partial x} \right| + w_z \left| \frac{\partial c\left(\mbx\right)}{\partial z} \right|
\end{equation}
For all images in this study, $w_x=$5 and $w_z=$1. The higher weighting factor for $w_x$ applies greater regularization in the lateral direction compared to the axial direction, accounting for the difference in lateral and axial k-space interrogation of ultrasound imaging systems. The regularization weight $\alpha$ is scaled by the grid spacing, [$\Delta x$, $\Delta z$], to maintain consistent regularization strength across different grid resolutions.

Initialization of the differentiable beamformer requires an \emph{a priori} velocity. To identify an optimal velocity to initialize the model, the first iteration of the differentiable beamformer was implemented using \emph{a priori} velocities between 1400\,m/s and 1600\,m/s. The CMPE was computed as a function of \emph{a priori} velocity and the velocity that minimized the CMPE was selected as the initialization velocity. This initialization velocity also provided an optimal global velocity for comparison to conventional beamforming. For these studies, a fixed number of 100 optimization iterations were performed, except for the phantom in Fig~\ref{fig:impact}, which used 200 optimization iterations.

\subsection{Grid Parameters}
Autofocusing was implemented using Cartesian grids for B-mode image reconstruction, CMPE evaluation, and velocity representation. Grid size and spacings for the experiments described in the following sections are detailed in Tab.~\ref{tab:grid_definitions}. The grid size and extent depended on the imaging field of view (e.g. for the rats, the anatomical depth of the rats varied from 16 to 26\,mm), while the grid spacing was determined based on resolution for the B-mode image and workstation memory limitations for the velocity and CMPE computation. For the computation of CMPE at each grid location in Tab.~\ref{tab:grid_definitions}, a 5$\times$5 kernel with spacing [$\Delta x$, $\Delta z$] = [$2\lambda/5$, $2\lambda/5$] centered on the grid location was used for all lag-1 common midpoint pairs.

\begin{table}[htbp]
    \centering
    \caption{Grid parameters for simulations, phantoms, and in vivo experiments.}
    \label{tab:grid_definitions}
    \begin{tabular}{llcc}
    \toprule
    \textbf{Target} & \textbf{Grid Type} & \textbf{Grid Extent} & \textbf{Grid Spacing} \\
                    &                    & \textbf{(mm)} & \textbf{[$\Delta x$, $\Delta z$] (mm)} \\
    \midrule
    \multirow{3}{*}{Rats~\cite{telichko2022noninvasive}} 
      & B-mode       & $[-16,16] \times [0,\textrm{Var.}]$     & [$\lambda/3$, $\lambda/3$] \\
      & Velocity     & $[-16,16] \times [0,\textrm{Var.}]$     & [1, 1] \\
      & CMPE         & $[-16,16] \times [3,\textrm{Var.}]$     & [1.50, 1.50] \\
    \midrule
    \multirow{3}{*}{IMPACT~\cite{ali2023sound}} 
      & B-mode       & $[-24,24] \times [0,45]$     & [$\lambda/2$, $\lambda/2$] \\
      & Velocity     & $[-24,24] \times [0,45]$     & [2, 2] \\
      & CMPE         & $[-24,24] \times [3,48]$     & [1.50, 1.41] \\
    \midrule
    \multirow{3}{*}{Inclusion~\cite{simson2023dbua}} 
      & B-mode       & $[-20,20] \times [1,40]$     & [$\lambda/3$, $\lambda/3$] \\
      & Velocity     & $[-20,20] \times [1,40]$     & [1, 1] \\
      & CMPE         & $[-20,20] \times [4,43]$     & [1.25, 1.22] \\
    \midrule
    \multirow{3}{*}{Phantoms~\cite{ali2021local}} 
      & B-mode       & $[-20,20] \times [0,40]$     & [$\lambda/2$, $\lambda/2$] \\
      & Velocity     & $[-20,20] \times [0,40]$     & [2, 2] \\
      & CMPE         & $[-20,20] \times [3,43]$     & [1.25, 1.25] \\
    \bottomrule
    \end{tabular}
\end{table}

\subsection{Simulated Experiments}\label{sec:ksims}
Simulated data was used to evaluate the merit of CMPE as a image focus metric. Simulated radiofrequency (RF) signals were generated with k-Wave \cite{treeby2010k} using a phantom with an isoechoic circular inclusion having an acoustic velocity of 1570\,m/s in a background medium having an acoustic velocity of 1540\,m/s. An isoechoic inclusion was created by scaling the mean density inversely proportional to the velocity to avoid specular reflections at the boundary. Diffuse scattering was achieved using sub-wavelength scatterers with random density and velocity changes within 5\% of the mean density and mean acoustic velocity. Plane wave transmissions were simulated at angles of $-28.5^\circ$ to $28.5^\circ$ with $0.5^\circ$ increments. The simulated probe consisted of 128 elements with 0.3\,mm pitch. The transmitted pulse had a center frequency of 4.8\,MHz, 100\% fractional bandwidth, and was a one-cycle Gaussian-weighted sinusoid. All transmits were focused in elevation at 20\,mm depth with a fixed lens. The simulated 3D domain had dimensions of 60\,mm axial $\times$ 51\,mm lateral $\times$ 7.4\,mm elevation. RF channel signals were captured and passed through a bandpass filter using the same spectra as the transducer. The RF channel data was then converted to the FSA basis using the REFoCUS method \cite{bottenus2017recovery,ali2019extending,lambert2020distortion} and then converted to IQ format.

\subsection{Phantom Experiments}\label{sec:phantom}
A publicly available dataset of a gelatin/silica phantom with heterogeneous acoustic velocity \cite{ali2023sound} was obtained to evaluate autofocusing with CMPE. The phantom is comprised of high-velocity alcohol-silica-gelatin cylindrical inclusions near the surface to induce aberration relative to the lower-velocity background material composed of silica-gelatin. Hyperechoic cylindrical inclusions of the same velocity as the background material were included in the deeper part of the phantom to serve as contrast targets and two rows of copper wires (127~$\mu$m in diameter) were placed in the middle of the phantom to serve as lateral resolution targets. Ground truth velocities of the high-velocity inclusions and background material were not available. RF channel data was collected from the phantom using a Vantage 256 system (Verasonics Inc., Kirkland, WA, USA) with a Philips (Amsterdam, Netherlands) L12-5 50\,mm transducer (256 elements, 0.2\,mm pitch) using a 6\,MHz transmit frequency. A Hadamard sequences was used on transmission and the recorded RF data was converted to the FSA basis and IQ format. Further details of this data and experiment can be found in \cite{ali2023sound}.

For evaluation of velocity estimation, RF channel data was obtained from four phantoms with calibrated velocity \cite{ali2021local}. The phantoms were an ATS 549 (Sun Nuclear, Melbourne, FL, USA; calibrated velocity of 1460\,m/s) and three graphite-gelatin phantoms (phantoms 1--3 with calibrated velocities of 1490, 1516, and 1540\,m/s). Chicken breast (calibrated velocity of 1575\,m/s) from a local supermarket was placed on top of the graphite-gelatin phantoms to create bi-layer phantoms. The RF channel data were collected using the Vantage 256 system with a Verasonics L12-3v transducer (192 elements, 0.2\,mm pitch, 7.8\,MHz transmit frequency) and a Hadamard sequence on transmission, and was then converted to the FSA basis and IQ format.

\subsection{In Vivo Experiments}
A selection of RF channel data from an obese Zucker rat model of metabolic-dysfunction associated steatotic liver disease from Telichko et al. \cite{telichko2022noninvasive} were used to evaluate the proposed method.  In this study, the obese Zucker rats were fed a high-fat diet to induce steatosis, which resulted in velocity variations in the liver. RF channel data was selected from rats 10, 11, and 18 from this study, which was collected from a Vantage 256 scanner and L12-3v transducer using a Hadamard sequence and a transmit frequency of 7.8\,MHz. The RF data was converted to the FSA basis and IQ format for use in the differentiable beamforming model.

\section{Results}
\begin{figure*}[t]
    \centering
    \includegraphics[width=\linewidth]{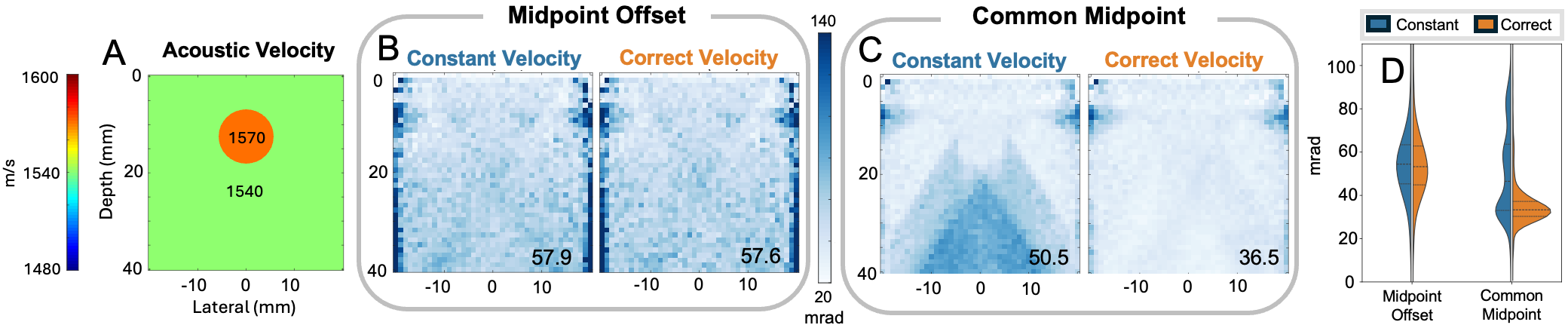}
    \caption{(A) The true speed of sound for an \emph{in silico} inclusion phantom used to simulate pulse-echo measurements for phase error evaluation. A comparison of phase error for a constant velocity reconstruction and the true velocity reconstruction is performed using (B) phase error with a lag-1 midpoint offset and (C) a common midpoint phase error (CMPE). (D) A comparison of the phase error distributions in the image between correct and incorrect velocity models for the lag-1 midpoint offset and common midpoint bases.}
    \label{fig:phase_measurement}
\end{figure*}

\subsection{Common Midpoint Phase Error as a Focusing Measure}
A demonstration of the ability of CMPE to identify and localize aberration is presented in a simulation example in Fig.~\ref{fig:phase_measurement}, with the simulated velocity field shown in Fig.~\ref{fig:phase_measurement}A. Fig.~\ref{fig:phase_measurement}B shows an image of the average phase error between $P_{i, j}(\mbx)$ and $P_{i+1, j+1}(\mbx)$, corresponding to subaperture pairs with a midpoint offset of 1 element (also called a lag-1 midpoint offset) for every spatial location in the medium; these aperture pairs do not share a common midpoint. Note that this phase error measurement shares similar qualities to the nearest-neighbor phase shift described in \cite{flax1988phase} where the lag-1 occurs between neighboring elements in the receive aperture. Whether focusing with an incorrect assumption of constant velocity or with the correct velocity field, the phase error images in Fig.~\ref{fig:phase_measurement}B do not qualitatively delineate the phase error. The source of the underlying phase error is ambiguous; the phase error due to natural signal decorrelation from the midpoint shift between signals cannot be distinguished from the phase error due to the incorrect velocity assumption, making lag-1 midpoint offset phase error a poor measure of phase aberration.

In Fig.~\ref{fig:phase_measurement}C, the CMPE is computed over all subaperture pairs sharing a lag-1 common midpoint, e.g. $P_{i,j}(\mbx)$ and $P_{i+1, j-1}(\mbx)$ for every spatial location in the medium. According to the VCZ theorem \cite{walker1997speckle}, the CMPE should have an expected value of 0\,mrad meaning that all error will be a result of an incorrect velocity assumption or propagation model. In Fig.~\ref{fig:phase_measurement}C, the incorrect constant velocity assumption shows increased CMPE beneath the circular lesion, corresponding to the magnitude of the phase aberration induced by a mismatch between the assumption of a constant reconstruction velocity and the true velocity field. Ultrasound image quality (e.g. resolution and contrast) is therefore expected to be worse in the darker regions of Fig.~\ref{fig:phase_measurement}C with high CMPE. With the correct reconstruction velocity field used for focusing, the CMPE below the lesion decreases and the entire region shows a low and homogeneous CMPE, indicating an absence of phase aberration and that improved image focusing is applied. The remaining non-zero phase error sources include errors from REFoCUS when converting the plane wave transmissions to FSA (especially the stronger errors at the edges at 8\,mm depth) and phase error from the straight-ray model for wave propagation.

The information content of the two phase error measures is evaluated by comparing their distributions before and after focusing in Fig.~\ref{fig:phase_measurement}D, with mean errors annotated with the center dotted line of the respective distribution. The mean midpoint offset phase error has a negligible change between incorrectly and correctly focused measurements (57.9\,mrad to 57.6\,mrad, $p=$0.72). The mean CMPE is significantly improved (50.6\,mrad to 36.5\,\,mrad, $p < $0.001) when a correct velocity model is applied for focusing. Fig.~\ref{fig:phase_measurement}D further shows that the distribution of midpoint offset phase error is nearly identical for constant and correct velocity focusing, as quantified by a Kullback-Leibler (KL) divergence of 0.035. Contrarily, CMPE displays a noticeable shift in distribution between the two focusing models, with a KL divergence of 1.16, showing that CMPE contains substantial phase aberration information not elucidated by the midpoint offset phase error.

\subsection{Autofocusing with Optimized CMPE}
\begin{figure*}[t]
    \centering
    \includegraphics[width=1.00\linewidth]{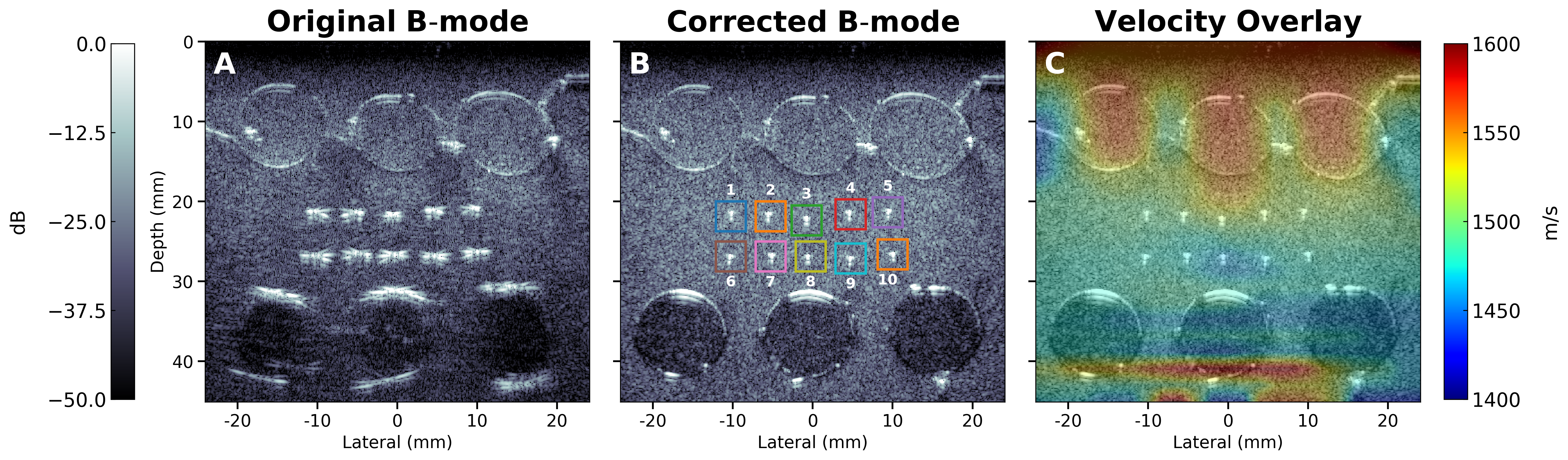}
    \caption{(A) The B-mode image of a tissue-mimicking phantom from Ali et al.~\cite{ali2023sound} using a constant velocity  (1518\,m/s) reconstruction model. Both rows of cylindrical inclusions display geometric distortions. Below the top row of isoechoic inclusions, the image is severely aberrated. The wire targets are poorly resolved with energy spread over 5\,mm laterally. The menisci on the lower row of inclusions are poorly focused. (B) The resulting B-mode image after CMPE autofocusing. The image appears brighter and more homogeneous, the menisci are focused well, and the geometrical distortion of the cylindrical inclusions is reduced. (C) Estimated velocity field overlaid on the corrected B-mode image.}
    \label{fig:impact}
\end{figure*}

Fig.~\ref{fig:impact}A shows the B-mode image reconstructed using an optimized global velocity of 1518\,m/s. Fig.~\ref{fig:impact}B shows the autofocused image of the same data, which displays hallmarks of improved focusing including tighter wire targets, brighter speckle, and tight and coherent specular menisci on the lower inclusions. The mean contrast and contrast to noise ratio (CNR) of the lower three cylindrical targets improved by -3.88$\pm$0.86\,dB and 0.41$\pm$0.15, respectively, between the two images. The mean CMPE improves from 171\,mrad before autofocusing to 81\,mrad after CMPE autofocusing, corroborating the qualitative improvements. The optimized velocity field for this model is overlaid on the B-mode image in Fig.~\ref{fig:impact}C, where higher velocity regions are shown coincident with the expected high velocity cylindrical inclusions. The remaining sections of the phantom are somewhat homogeneous with a few artifacts in velocity below the lower center cylinder. A video demonstrating the autofocusing of the B-mode image along with the corresponding estimated velocity field and velocity field overlaid on the B-mode image is provided in the supplementary material. The video shows the B-mode image and velocity estimates as the model updates through its iterative process, where it can be observed that localized distortions in the B-mode image are corrected within the early iterations (e.g. the wire targets sharpen) and more global distortions are corrected in the later iterations (i.e. the image slowly expands). This is coincident with the general structure of the velocity field identified in the early iterations and the global velocity increasing or decreasing in the later iterations.

\begin{table}[ht]
    \centering
    \caption{Lateral FWHM Measurements Across ROIs}
    \begin{tabular}{lccccc}
        \toprule
        & ROI 1 & ROI 2 & ROI 3 & ROI 4 & ROI 5 \\
        \midrule
        Aberrated FWHM (µm) & 1821 & 310  & 1310 & 524  & 929  \\
        Corrected FWHM (µm) & 179  & 179  & 202  & 167  & 190  \\
        \bottomrule
    \end{tabular}
    
    \vspace{0.5em}

    \begin{tabular}{lccccc}
        \toprule
        & ROI 6 & ROI 7 & ROI 8 & ROI 9 & ROI 10 \\
        \midrule
        Aberrated FWHM (µm) & 1869 & 1417 & 2012 & 1071 & 1250 \\
        Corrected FWHM (µm) & 190  & 167  & 167  & 179  & 202  \\
        \bottomrule
    \end{tabular}
    \label{tab:fwhm}
\end{table}

Table~\ref{tab:fwhm} provides the lateral full-width half-maximum (FWHM) of the wire targets from Fig.~\ref{fig:impact}A and B. The average reduction in FWHM of the top row of wires is 795\,\textmu m, while the bottom row improved by 1343\,\textmu m on average. After CMPE autofocusing, the average FWHM over all wire targets is 182\,\textmu m, which is close to the theoretical diffraction limit of 148\,\textmu m (based on an assumed constant velocity of 1540\,m/s).

\begin{figure*}[!htbp]
    \centering
    \includegraphics[width=1.00\linewidth]{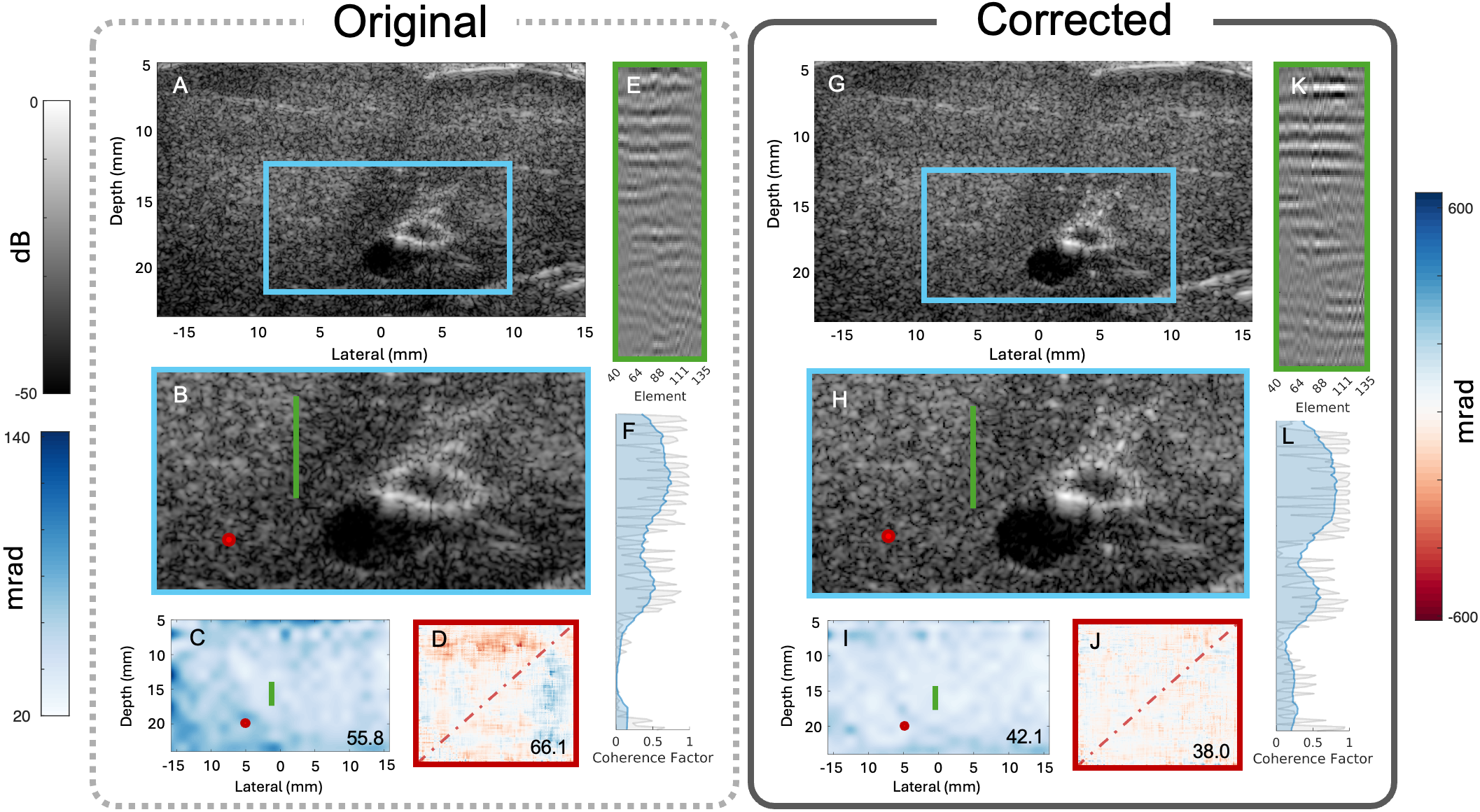}
    \caption{\emph{In vivo} example of CMPE autofocusing from rat 10. The original B-mode images (left) are beamformed with a constant acoustic velocity, and the proposed images (right) utilize CMPE autofocusing. (A) \& (G) B-mode images reconstructed using the two methods, with the regions marked by the blue boxes magnified in (B) \& (H). Images of the CMPE are shown in (C) and (I) with the spatial mean of the CMPE displayed in the lower right in mrad. (D) \& (J) Phase shifts between beamformed samples for all active transmit and receive sub-aperture-pairs for the location marked by the red dots in (B) and (H), respectively. The phase shifts above and below the dotted line should be symmetric due to acoustic reciprocity. The proposed method leads to an overall lower mean CMPE value (in mrad, lower right) for this location. (E) \& (K) RF channel data at the location of the green line in (B) and (H), respectively, with applied geometric focal delays. (F) \& (L) The coherence factor (CF) of the RF channel data in (E) \& (K) as a function of depth (gray) and with a 20-sample moving average (blue).}
    \label{fig:focus}
\end{figure*}

Fig.~\ref{fig:focus} shows \emph{in vivo} images of the liver including the portal vein (the hyperechoic ring) and the hepatic vein from rat 10. The original images are reconstructed using a constant velocity (optimized to be 1568 m/s) and shown in Fig.~\ref{fig:focus} on the left (Original). The proposed CMPE autofocused image is shown on the right (Corrected). Figs.~\ref{fig:focus}A and \ref{fig:focus}G compare B-mode images before and after CMPE autofocusing. The CMPE autofocused image displays an overall brighter speckle and markedly tighter point features, and the contrast and CNR of the hepatic vein improved by -2.64\,dB and 0.34, respectively. Figs.~\ref{fig:focus}B and \ref{fig:focus}H show magnified views of the region outlined by the blue bounding box, highlighting the hepatic vein and surrounding liver structures. The CMPE autofocused image shows vessels with higher contrast, tighter boundaries, and improved specular reflector definition.

Figs.~\ref{fig:focus}C and \ref{fig:focus}I show the spatial CMPE images throughout the field of view, where CMPE autofocusing is observed to reduce the CMPE, particularly near the top- and bottom-left of the image. The speckle in these regions increases in brightness, indicating improved focusing. Quantitatively, the mean CMPE is reduced from 55.8 to 42.1\,mrad after CMPE autofocusing. Figs.~\ref{fig:focus}D and \ref{fig:focus}J compare the signed phase shifts between pairs of common midpoint apertures that contribute to the mean CMPE, measured at the red dot, before and after CMPE autofocusing. CMPE autofocusing reduces the magnitude of the phase shift, indicated by a whiter appearance.

Figs.~\ref{fig:focus}E and \ref{fig:focus}K show the received channel data (i.e, axial sampling of the focused echo-wavefronts) corresponding to the green lines in the respective images. The CMPE autofocused channel data have a higher amplitude and more coherent wavefronts as a result of improved focusing. The improved coherence of the wavefronts is confirmed in Figs.~\ref{fig:focus}F and \ref{fig:focus}L, where the CF of the received channel data is plotted vertically in gray over the axial extent of the measurement. A 20-sample moving average is overlaid in light blue. After CMPE autofocusing, the CF is increased in Fig.~\ref{fig:focus}L compared to Fig.~\ref{fig:focus}F.

\begin{figure*}[t]
    \centering
    \includegraphics[width=1.0\linewidth]{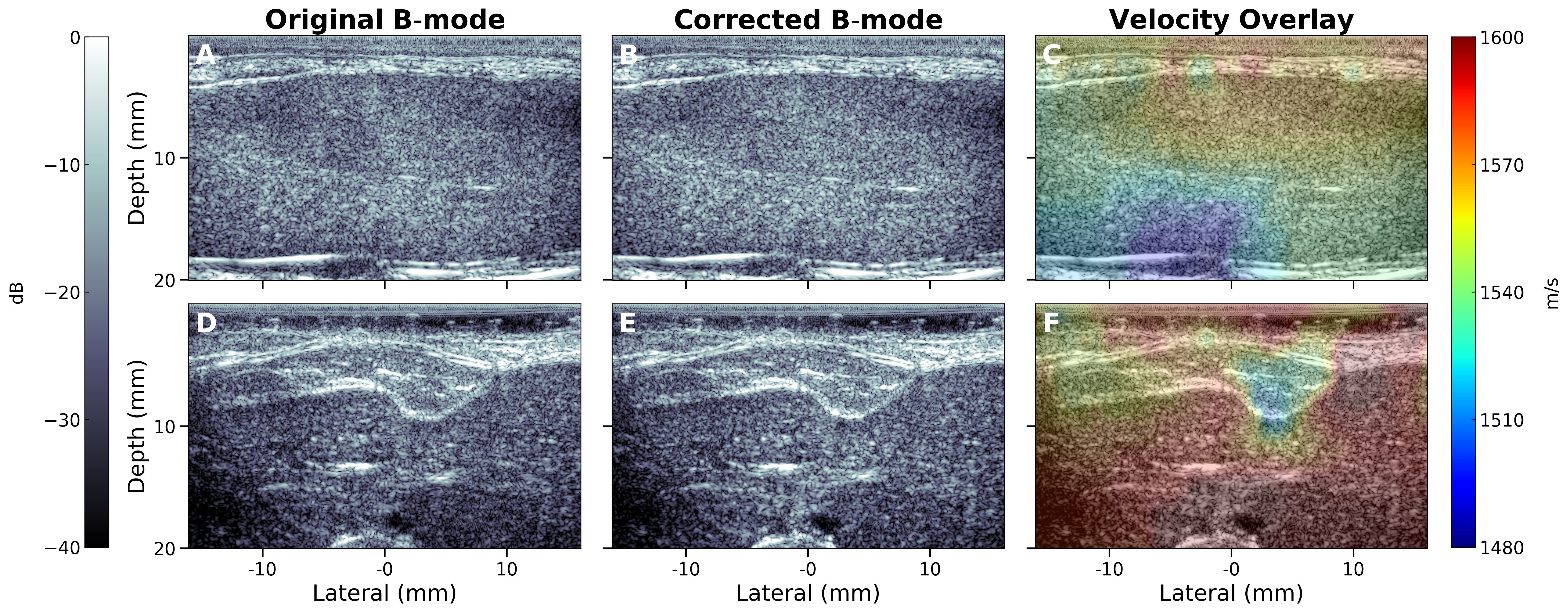}
    \caption{(A) \& (D) B-mode images of the livers in rats 18 and 11, respectively, before CMPE autofocusing. The models were initialized using an optimized constant velocity (1546.0 and 1570.0 m/s, respectively). (B) \& (E) B-mode images of the same livers after CMPE autofocusing. (C) \& (F) show the velocity estimates overlaid on the CMPE autofocused B-mode image. While the velocity estimates are used to parameterize the beamforming delays to optimally focus the B-mode image, they can also potentially be used as a biomarker for disease. The values here show some spatial agreement with anatomical tissue and are quantitatively plausible, with velocity values corresponding to the literature for these tissues.} 
    \label{fig:qualitative}
\end{figure*}

Fig.~\ref{fig:qualitative} shows additional uncorrected and CMPE autofocused (corrected) images from two rats, with the estimated velocity field overlaid on the corrected B-mode image. The image in Fig.~\ref{fig:qualitative}A, beamformed with an optimized global average velocity of 1546\,m/s, shows two liver lobes layered vertically and delineated by a horizontal interface in the center of the B-mode image. The corrected B-mode image (Fig.~\ref{fig:qualitative}A) shows a similar appearance to the uncorrected image with small improvements, such as a more homogeneous brightness in the liver and stronger delineation between the two lobes. The velocity estimates show regions with different velocity that transition smoothly in the axial and lateral directions between the different velocity regions. The mean velocities of the top and bottom regions are 1559\,m/s and 1521\,m/s, respectively, and are relatively close to the global optimized velocity.

Fig.~\ref{fig:qualitative}D also shows distinct liver lobes in a rat liver, identified by a brighter liver lobe proximal to the abdominal tissue and a darker, larger lobe distal to the abdominal tissue. Fig.~\ref{fig:qualitative}E shows more significant changes in the image due to CMPE autofocusing, where the speckle brightness in the lower liver is brighter and the visualization of the hepatic vein (located at 18\,mm depth and 0--3\,mm laterally) improves significantly. The contrast and CNR of the hepatic vein here improved by -4.87\,dB and 0.28, respectively. In the velocity overlay in Fig.~\ref{fig:qualitative}F, the tip of the median lobe presents with a low velocity of approximately 1537\,m/s and is positioned directly over the hepatic vein, which is located in a region of tissue with a velocity closer to 1600\,m/s. The velocity regions surrounding the low velocity region transition more sharply to higher velocities compared to what was observed in Fig.~\ref{fig:qualitative}C. The CMPE was reduced from 63.9\,mrad to 45.2\,mrad in Fig.~\ref{fig:qualitative}A--B and from 110.2\,mrad to 86.3\,mrad in Figs.~\ref{fig:qualitative}D--E.

Supplemental videos are provided to show CMPE autofocusing for the images of the rats in Figs.~\ref{fig:focus} and~\ref{fig:qualitative}. Two additional videos for each rat from slightly different views are also provided. These videos more distinctly show the changes in image quality while simultaneously displaying the velocity field updates and an overlay of the velocity field on the B-mode image. For rat 11, one of the views shows the transducer rotated 90$\degree$ to the position used in Figs.~\ref{fig:qualitative}D--F. Similar to the phantom video, localized distortions in the B-mode image are corrected in the early iterations and global distortions are corrected in later iterations.

\subsection{Velocity Reconstruction with CMPE Minimization}
\begin{figure*}[!htbp]
    \centering
    \includegraphics[width=1.00\linewidth]{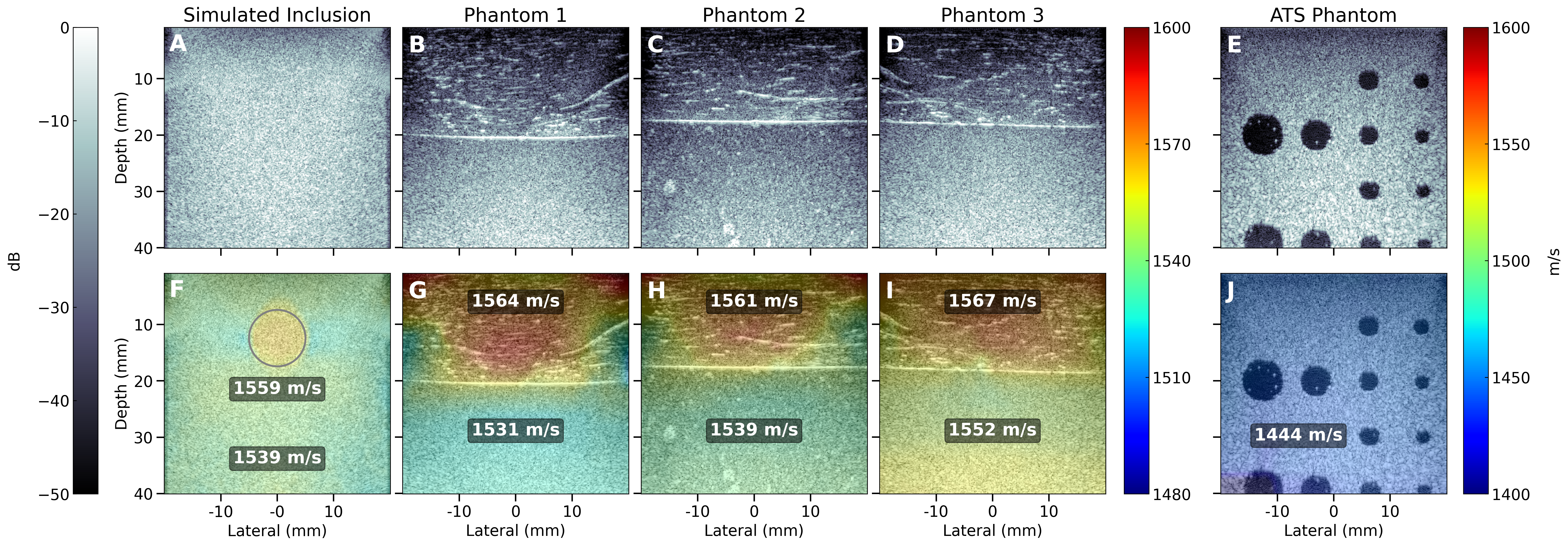}
    \caption{Velocity estimation in simulation (A, F) and phantom examples (B--E, G--J) from CMPE autofocusing. The corrected B-mode image from CMPE autofocusing is shown for (A) the simulated isoechoic inclusion from Fig.~\ref{fig:phase_measurement}A, (B-D) bi-layer phantoms with boundaries defined at 20, 17, and 18\,mm for phantoms 1--3, respectively, and (E) a homogeneous velocity ATS 549 phantom. (F--J) Estimated velocities are overlaid on the corrected B-mode images corresponding to (A--E). The mean velocity estimated in the simulation inclusion, simulation background, phantom regions, and chicken layers are displayed on the images.}
    \label{fig:phantoms}
\end{figure*}

Velocity estimates from media with calibrated velocities are shown in Fig.~\ref{fig:phantoms} to demonstrate the accuracy of the velocity estimates. The corrected B-mode images are shown in the top row and the estimation of the velocity field is overlaid on the corrected B-mode image in the bottom row. In the simulated isoechoic lesion (A \& F), the estimated velocity of the background medium is close to the ground truth value (bias of -1\,m/s), while the estimated velocity in the inclusion is 11\,m/s below the ground truth. A similar sized bias is observed in the chicken layer of the bi-layer phantoms (B--D, G--I) and the ATS phantom (E \& J), ranging from 8--14\,m/s below the calibrated value. The velocity estimation in the graphite-gelatin phantoms show a greater positive bias, with errors ranging from 12\,m/s below to 42\,m/s above the calibrated velocities.

\section{Discussion}
To measure focusing quality, we have introduced the CMPE as a reliable and robust measure applicable to diffuse scattering in heterogeneous media. Most focusing criteria are derived from the coherence of the wavefield across the receive aperture. Receive aperture coherence has been widely used to measure focusing quality and aberration in speckle, either via its phase shift \cite{flax1988phase}, coherent sum magnitude \cite{nock1989phase}, or other related quantities \cite{mallart1994adaptive}. However, the diffuse sub-resolution scatterers encountered in medical ultrasound imaging produce speckle, which exhibits only partial coherence across the receive aperture \cite{mallart1991van}. Speckle signals across the receive aperture increasingly decorrelate as their spacing increases at a rate predicted by the VCZ theorem (\eqref{eq:translatingapertures}, when $\Delta\mbx_t=0$ but $\Delta\mbx_r\ne 0$).  Speckle decorrelation manifests as phase shifts that are fundamentally indistinguishable from phase shifts caused by aberration, introducing uncertainty in phase shift estimation as quantified in \eqref{eq:jitter}. Fig.~\ref{fig:phase_measurement}B demonstrates that this uncertainty can overwhelm the phase errors from aberration. For example, although ideal focusing is achieved under the correct velocity, there is little or no change in the measured phase errors (57.6\,mrad) compared to the incorrect velocity model (57.9\,mrad), meaning that the measured phase errors in Fig.~\ref{fig:phase_measurement}B are largely a product of speckle decorrelation and not aberration.

We have also demonstrated that the CMPE is an excellent objective function when using a differentiable beamforming model to achieve autofocusing. Focusing measures based on receive aperture coherence are traditionally not maximized by correct focusing in diffusely scattering media, making them poor optimization objectives. For instance, one can artificially exceed the $\textrm{CF}=\frac{2}{3}$ expected of correct focusing in speckle \cite{mallart1994adaptive} by overfitting to the jitter, thereby introducing artificial aberration into the focusing to produce false point targets. In contrast, common midpoint images of speckle are very highly correlated and have minimal jitter (combining \eqref{eq:gamma12_cmp} and \eqref{eq:jitter}), allowing phase errors to be attributed almost entirely to aberration (Fig.~\ref{fig:phase_measurement}). 
Prior phase aberration correction methods utilized an estimate of the isoplanatic patch size when applying correction \cite{ali2023aberration,chau2019locally,lambert2020distortion}. The isoplanatic patch size determines the rate at which the phase errors change with respect to translation of the beamformer's focal point. Reconstruction quality in these models is sensitive to the selected isoplanatic patch size, which acts as a regularizer to constrain the large solution space of phase screens. However, we note that the actual isoplanatic patch size is a physical consequence of acoustic velocity variations. Our model instead parameterizes the beamformer as a function of the underlying speed of sound, such that estimating the isoplanatic patch size becomes unnecessary because it is implicitly enforced by the geometry of the velocity field and the regularization applied to the velocity field. This parameterization naturally constrains the solution space of spatially-varying aberration laws to smoother variations compatible with the physical model of medium velocity, and is observed to be capable of producing high focusing quality at all image locations.  Note that implicit regularization also occurs by the representation of the velocity field on a coarser grid than the imaging grid, which also effects the applied isoplanatic patch.

A byproduct of CMPE autofocusing differentiable beamforming is the estimate of the underlying acoustic velocity. While these velocity fields demonstrate values consistent with soft tissue and may appear to spatially correlate with anatomical features visible in the B-mode images, it is important to note that these velocity estimates are optimized to minimize CMPE and improve focusing and not explicitly to faithfully model the underlying velocity field (see velocity estimates in Fig.~\ref{fig:phantoms}). The reflection tomographic velocity estimation problem is notoriously ill-posed and multiple velocity fields can lead to similar focus quality \cite{biondi20063d}. The limited aperture in pulse-echo imaging leads to insufficient angular sampling, leading to a challenging reconstruction problem for internal volume states as suggested by Kirchhoff's Integral Theorem~\cite{schneider1978integral}. Nevertheless, preliminary work \textit{in silico}~\cite{simson2023dbua} suggests that CMPE autofocusing can lead to spatial and quantitative velocity estimates that can aid in the interpretation of the corrections that occur in the B-mode images. For example, Fig.~\ref{fig:qualitative}B shows an example of an image with small corrections while Figs.~\ref{fig:impact}B and~\ref{fig:qualitative}E show examples of more dramatic corrections. The velocity field corresponding to Fig.~\ref{fig:qualitative}B shows smoother and smaller variations in velocity, leading to more global corrections in the image rather than localized corrections (see supplementary video rat18\_view02\_S8), while the velocity fields for Figs.~\ref{fig:impact}B and~\ref{fig:qualitative}E show sharper transitions and larger variations in velocity across the lateral dimension, leading to greater improvements in localized distortions, specifically in the targets directly below the sharper velocity transitions. The greater improvement in Fig.~\ref{fig:qualitative}E compared to Fig.~\ref{fig:qualitative}B is due to larger phase differences induced by the larger velocity variations as exemplified in Fig.~\ref{fig:phase_measurement} where the CMPE reduction is more substantial in the region directly beneath the aberrating inclusion.

For these velocity fields, the reader is cautioned on interpreting velocities in the first approximately 3\,mm of the images where the common midpoint model breaks down due to diffraction, lens effects, and selection of the subaperture size. Velocities can be estimated in this region due to the model design, but will accumulate error because the CMPE in this region is unable to influence the velocity estimates, and therefore the velocity estimates are not trustworthy. Similarly, the reduced number of ray-paths intersecting edges of the image and the fewer available common midpoints at the image boundaries result in fewer measurements available for CMPE calculation and optimization and produce less accurate velocity estimates.

There are several sources of potential errors that our beamforming model does not address. Our beamforming model employs the geometric straight-ray wave propagation model, ignoring any refraction or diffraction effects from heterogeneous wave velocity. These effects can manifest as spurious CMPE and may cause this model to break down in media with significant sound speed heterogeneity (e.g., transcranial imaging through the skull). At the same time, this specific error could be viewed as a confidence metric of the accuracy of a beamforming model, such as the straight-ray model used here. For example, the straight-ray model breaks down under strong diffracting elements or refracting layers, such as a checkerboard phantom of acoustic velocity~\cite{simson2023dbua}, can be identified by residual CMPE after CMPE autofocusing. This potentially explains why anatomical regions visible in the B-mode images do not always correlate with the velocity map and why estimated velocities deviate from ground truth or calibrated measurements. For example, in Fig.~\ref{fig:impact}, the regions of higher velocity corresponding to the isoechoic cylinders appear narrower and longer than would be assumed in the CMPE autofocused B-mode image, likely due to the straight-ray model deviating from the refracted paths through these targets. Models accounting for the refraction and diffraction, such as those used in Ali et al. \cite{ali2022distributed} or WEMVA~\cite{sava2004wave} may produce less CMPE and more accurate velocity estimation.

Similarly, thermal noise, reverberation noise (either diffuse or specular), or strong specular scatterers away from the focal point, $\mbx_f$, can manifest as spurious CMPE that is not removed by the optimization process. Thermal and diffuse reverberation noise is mitigated somewhat by the use of the correlation threshold on the common midpoint phase error, which avoids incorporating large phase errors that are more likely to be associated with these sources of noise. The correlation thresholding and subaperture beamforming over the 17 elements in this implementation may reduce the influence of all of these noise sources and improve SNR, although subaperture beamforming also introduces decorrelation between the lag-1 common midpoint signals due to the underlying aberration within the subapertures. Smaller apertures can may be used to reduce this decorrelation at the expense of worse SNR. In addition, a smaller subaperture increases the number of possible common midpoint phase computations, thereby increased computational effort. A more rigorous study of the subaperture size on these factors may better optimize the subaperture selection.

Historically, aberration correction methods cannot accurately measure and correct aberration in hypoechoic and anechoic regions and may require methods to identify anechoic regions such as in Chau et al. \cite{chau2019locally} or assuming a constant velocity over the anechoic region, as in Sanabria et al. \cite{sanabria2018spatial}. Avoiding the computation of CMPE in anechoic regions would mitigate erroneous gradients in the backpropagation step and improve reflection tomographic velocity reconstruction with differentiable beamforming.

Errors in aperture synthesis can manifest as CMPE, such as the effect of a limited physical aperture in virtual source models (e.g., plane or diverging wave synthetic transmit aperture), or inaccurate multistatic data recovery using spatially-coded apertures (e.g., the REFoCUS technique \cite{bottenus2017recovery,ali2019extending}). This can be observed, for example, in Fig.~\ref{fig:phase_measurement}C in the phase errors at the edges at approximately 8\,mm depth, which are due to errors in REFoCUS when converting from the plane wave basis to the FSA basis. While the plane wave or Hadamard sequences used to capture the data in this study are not required for CMPE autofocusing because REFoCUS can be used to recover the multistatic data, the choice of pulse sequence will influence the amount of residual CMPE unrelated to aberration. Finally, the CMPE was derived under the assumption of an incoherent source in a diffusely scattering media \eqref{eq:spatial_incoherence}, and not for coherent sources like specular reflectors, although prior work suggests that the mutual coherence of a coherent source exhibits similarly ideal CMPE \cite{goodman2015statistical,rachlin1990direct}. Collectively, these sources of potential errors in CMPE manifested here as non-zero CMPE after optimization or when using the ground truth velocity, but did not prevent the optimization model from achieving autofocusing.

The primary computational bottlenecks of CMPE autofocusing are the iterative velocity updates and the gradient calculations through automatic differentiation. Using our current unoptimized code, processing a single image frame requires approximately 2--3 minutes on a desktop workstation with an NVIDIA RTX 3080 GPU. However, we note several changes that could lead to real-time implementation: (1) because the local distortions were fixed within approximately the first 20 iterations, we could limit the optimizer to CMPE autofocusing over fewer iterations, (2) further tuning of the learning rate could enable faster convergence (learning rate was only briefly empirically explored), and (3) the use of dedicated hardware or software beamformers to perform more efficient auto-differentiation or faster GPUs could be employed.

\section{Conclusions}
We have introduced an autofocusing method based on common-midpoint phase error (CMPE) optimization and differentiable beamforming. This approach models the beamforming process as a straight-ray model of wave propagation through heterogeneous media and uses the CMPE in a loss function to update the velocity field of the model. The autofocusing method corrects for aberrations due to a constant velocity assumption for beamforming and, as a byproduct, produces an estimate of the velocity field. Substantial improvements in ultrasound image focusing were observed in phantoms with heterogeneous velocity and was observed to improve focusing \emph{in vivo} in rat livers. The resulting CMPE and velocity field can be used to aid in the interpretation of the aberrations.

\section*{Acknowledgments}
This work was supported in part by the National Institute of Biomedical Imaging and Bioengineering under Grants R01-EB027100 and K99-EB032230.
\bibliography{abbrev,references}
\end{document}